\documentclass[summary]{arxiv-ursi}

\usepackage{tumcolors}
\usepackage{tumlang}
\usepackage{tummath}
\usepackage{cite}
\usepackage{pgfplots}
\usepackage{pgfplotstable}
\usepgfplotslibrary{units}
\usetikzlibrary{3d}
\pgfplotsset{compat=newest}
\usepackage{amsfonts}
\usepackage{amssymb}

\renewcommand{\imu}{\i}
\DeclareMathOperator{\traceu}{Tr}
\DeclareSIUnit{\arbitraryunit}{a.u.}

\usetikzlibrary{arrows}
\pgfplotsset{compat=newest}

\title{Reducing the reflection error of PML absorbing boundary conditions within a generalized Maxwell-Bloch framework}

\author{Johannes Popp*, Lukas Seitner, Michael Haider, and Christian Jirauschek}

\affiliation{%
  {Department of Electrical and Computer Engineering, Technical University of Munich,\\
    Arcisstr. 21, 80333 Munich, Germany}
}

\begin{document}

\maketitle

\begin{abstract}
  We demonstrate a full-wave numerical Maxwell-Bloch simulation tool including perfectly matched layer (PML) absorbing boundary conditions. To avoid detrimental reflection errors at the boundary of the simulation domain, an adapted PML model is introduced, which takes into account impedance mismatch effects arising from the internal quantum system. For the numerical validation of the modified PML model the simulation tool is applied to the active gain medium of a terahertz quantum cascade laser (QCL) structure. Improved absorbing characteristics for the truncation of active gain media in our Maxwell-Bloch simulation approach are obtained.
\end{abstract}

\section{Introduction}

The Maxwell-Bloch equations are a valuable tool for modeling light-matter
interaction in nonlinear
optics~\cite{boyd2020nonlinear,jirauschek2019optoelectronic}. Here, the
evolution of a discrete-level quantum system is described by the Bloch
equations, which are coupled to Maxwell's equations and provide a classical
description of the optical field. In recent years, this equation system has
been applied to the simulation of various active photonic devices in order to
describe nonlinear optical
phenomena~\cite{slavcheva2002coupled,slavcheva2008model,menyuk2009self,talukder2014quantum,ziolkowski1995ultrafast,tzenov2016time,jirauschek2017self}.
The work presented here is based on the open-source solver tool
mbsolve~\cite{riesch2021mbsolve,Seitner_2021}, where the generalized
one-dimensional Maxwell-Bloch equations are treated without invoking the
rotating wave approximation (RWA).

In order to solve the Maxwell-Bloch equations, mbsolve uses the
finite-difference time-domain (FDTD) method, one of the standard approaches
in computational electrodynamics~\cite{taflove2005computational}. A main
challenge in simulating optical devices in open radiation problems is the
truncation of the FDTD lattice~\cite{taflove2005computational}. Here, the
main idea is to introduce a highly absorbing, reflectionless layer at the
outer boundary of the spatial FDTD grid. In 1994, Berenger introduced
perfectly matched layers as a non-physical
absorber~\cite{berenger1994perfectly}. His approach is based on the so called
field-splitting method, where the field components are split into two
orthogonal components resulting in modified Maxwell's equations. Based on
this work, Fang~\textit{et
  al.}~\cite{fang1995generalized,fang1996generalized} introduced a general PML
method for lossy materials by extending the original PML approach by
Berenger. A uniaxial anisotropic PML (UPML) absorber was introduced by
Gedney~\cite{gedney1996anisotropic2}, where the mathematical model of
field-splitting is replaced by a more physical model based on the Maxwellian
formulation. In a consecutive publication, he extended his approach for the
absorption of fields in lossy and dispersive
materials~\cite{gedney1996anisotropic1}. Wang~\textit{et
  al.}~\cite{wang2011implementation} presented a PML approach for the
truncation of a gain medium within a semiconductor Maxwell-Bloch framework
for both, un-split and field-splitting methods.

\begin{figure}
  \centering
  \input{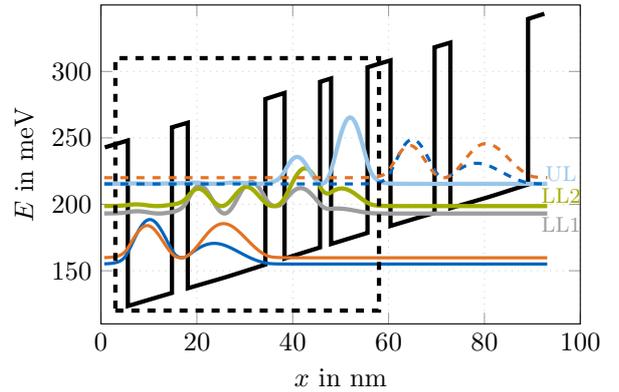}
  \caption{Calculated conduction band profile and probability densities of the investigated THz QCL structure~\cite{burghoff2014terahertz,tzenov2016time} with an optical transition frequency of \SI{3.5}{\tera\hertz}. The upper and lower laser levels are represented by bold solid lines. The dashed rectangle comprises a single QCL period.}
  \label{fig:wavefunctions}
\end{figure}

In this paper, we introduce the theoretical model for the integration of UPML
boundary conditions into the existing code base of mbsolve. For the
evaluation of the PML implementation we will use a well-studied quantum
cascade laser design~\cite{burghoff2014terahertz,tzenov2016time} as an
exemplary device. A QCL is a special type of semiconductor laser emitting in
the mid-infrared and THz frequency range. Its active gain medium consists of
a multiple quantum well heterostructure. Quantized electron states act as
laser levels and charge transport occurs through intersubband transitions
between the states~\cite{jirauschek2014modeling}. In
Fig.~\ref{fig:wavefunctions}, the bandstructure and calculated wavefunctions
of the investigated THz QCL gain medium are illustrated. Optical transitions
appear between the upper level UL and the two lower levels LL1 and LL2. All
relevant simulation parameters can be found in~\cite{Seitner_2021}.

\section{Simulation model}

The Maxwell-Bloch equations describe the light-matter interaction of the
optical field with a quantum mechanical system. The density operator
$\function{\op{\rho}}{t,x}$ contains all information about the quantized
electronic states. In order to minimize the workload for the numerical
simulations, we reduce the model to a single spatial coordinate. The validity
of this simplification is guaranteed as we restrict our investigations to
optoelectronic devices with widespread waveguides in propagation and lateral
direction, thus allowing the separation of the electromagnetic field into
transverse and longitudinal modes~\cite{jirauschek2019optoelectronic}. In the
following, the propagation direction is $x$ and the optical field is
represented by the field components $\function{E_z}{x,t}$ and
$\function{H_y}{x,t}$ with $z$ and $y$ being the transversal coordinates. The
density operator $\function{\op{\rho}}{t,x}$ is governed by the Liouville-von
Neumann master equation
\begin{equation}
  \partial_t{\op{\rho}}=-\imu\hbar^{-1}\commutator{\op{H_0}-\op{\mu}_z
    E_z}{\op{\rho}}+\function{\mathcal{D}}{\op{\rho}}\,,
\end{equation}
where $\hbar$ ist the reduced Planck constant, $\op{H_0}$ is the system
Hamiltonian, $\op{\mu}_z$ is the dipole moment operator, and
$\commutator{\cdot}{\cdot}$ denotes the commutator
$\commutator{\op{A}}{\op{B}}=\op{A}\op{B}-\op{B}\op{A}$. The dissipation
superoperator $\function{\mathcal{D}}{\op{\rho}}$ models the interaction with
the environment. From the dipole moment of the quantum system the macroscopic
polarization term $P_{z,\mathrm{qm}}$ is obtained by
\begin{equation}
  {P_{z,\mathrm{qm}}}=n_\mathrm{3D}\traceu\left\lbrace\op{\mu}_z {\op{\rho}}\right\rbrace\, ,
\end{equation}
where $n_\mathrm{3D}$ is the carrier number density.

The optical field propagation is described classically using Faraday's law
and Ampere's law for the electric field $E_z$ and magnetic field $H_y$. By
utilizing the auxiliary differential equation approach the update equations
in the UPML medium can be derived. Using the relation
$\function{D_z}{\omega}=\varepsilon_0\function{\varepsilon_\mathrm{r}}{\omega}\function{E_z}{\omega}$
and taking into account the constitutive parameter
$s_x=1+\frac{\sigma_x}{\imu\omega\varepsilon_0}$, the time evolution equation
for the electric flux density is given by
\begin{equation}
  \partial_t D_z=\partial_x H_y-\frac{\sigma_x}{\varepsilon_0} D_z\, ,
\end{equation}
with the UPML conductivity $\sigma_x$.
Furthermore, the time evolution of the electric field is represented by
\begin{equation}
  \begin{aligned}
    \partial_t E_z =~ &
    \left(\varepsilon_0\varepsilon_{\mathrm{r},\infty}\right)^{-1} \\ & \left(\partial_t D_z - \sigma_0 E_z - \partial_t P_{z, \mathrm{qm}} - \sum_i\partial_t P_{z, \mathrm{class}}^i \right),
  \end{aligned}
\end{equation}
where $\varepsilon_0\varepsilon_{\mathrm{r},\infty}$ is the product of the
vacuum permittivity and the relative permittivity in the infinite-frequency
limit, $\sigma_0$ is the material conductivity and $\sum_i\partial_t P_{z, \mathrm{class}}^i $ is the multi-polarization term accounting
for bulk and waveguide dispersion~\cite{taflove2005computational}. We also
take into account the macroscopic polarization of the quantum system in the
PML in order to reduce the reflection error at the interface layer between
the absorbing boundary and the main simulation region. Several generic models
such as e.g.\ Drude and Lorentz are implemented and can be combined. The time
evolution equation of the magnetic field can be written as
\begin{equation}
  \partial_t H_y = \mu^{-1}\partial_x E_z-\frac{\sigma_x}{\epsilon_0}H_y\, ,
\end{equation}
with the permeability $\mu=\mu_0\mu_r$. In order to reduce the parasitic
reflection errors from the PML layers, the conductivity $\sigma_x$ in the PML
layers is gradually increased along the propagation direction. Therefore, the
conductivity is varied using a smooth polynomial with depth $x$ in the PML
layer~\cite{taflove2005computational,gedney1996anisotropic2}
\begin{equation}
  \function{\sigma_x}{x}=\left(x/d\right)^m\sigma_{x,\mathrm{max}}\, .\label{eq:sigma_x}
\end{equation}
The optimal choice for $\sigma_{x,\mathrm{max}}$ is given
by~\cite{taflove2005computational,gedney1996anisotropic1}
\begin{equation}
  \sigma_{x,\mathrm{opt}}=\frac{0.8\left(m+1\right)}{\eta_0\Delta\sqrt{\varepsilon_\mathrm{r,eff}\mu_\mathrm{r,eff}}}\, ,
\end{equation}
where $\eta_0$ is the free-space wave impedance, $\Delta$ is the lattice-cell
dimension, and $\varepsilon_\mathrm{r,eff}$ and $\mu_\mathrm{r,eff}$ are
constants representing the effective relative permittivity and permeability,
respectively. The values of $\varepsilon_\mathrm{r,eff}$ and
$\mu_\mathrm{r,eff}$ should be chosen either to be mean values of the
physical parameters or the values at the wavenumber of the fundamental mode
in the waveguide~\cite{taflove2005computational}.

\section{Results}

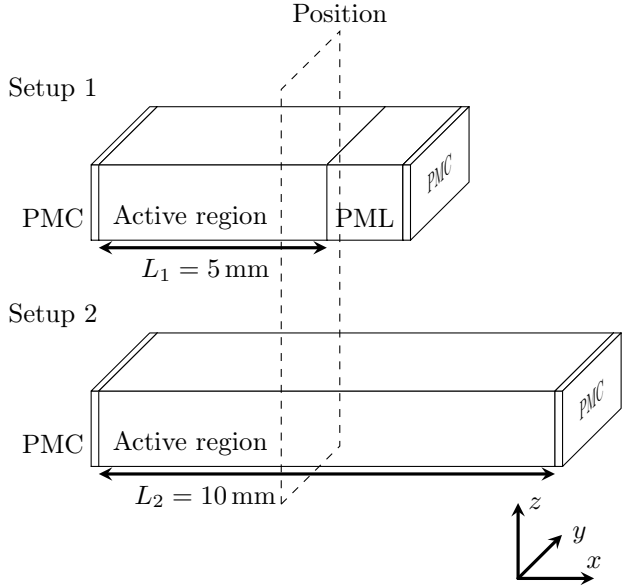
\begin{figure}[ptb]
  \centering
  \begin{tikzpicture}
  \tikzstyle{myline}=[black, line cap=round,line join=round];
  %
  % coordinate system
  \tikzstyle{cs}=[myline, very thick, ->, >=stealth];
  \tikzstyle{ld}=[myline, very thick, <->, >=stealth];

  \begin{scope}[shift={(6, 0, 5)}]
    \draw[cs] (0, 0, 0) -- (1, 0, 0) node[above] {$x$};
    \draw[cs] (0, 0, 0) -- (0, 1, 0) node[right] {$z$};
    \draw[cs] (0, 0, 0) -- (0, 0, -1.5) node[right] {$y$};
  \end{scope}
  %% Setup 2
  % PMC
  \tikzstyle{act}=[myline, fill=white];
  \begin{scope}[shift={(0, 1.1, 0)}]
    \coordinate (actA) at (0, 0, 4);
    \coordinate (actB) at (0.1, 0, 4);
    \coordinate (actC) at (0.1, 1, 4);
    \coordinate (actD) at (0, 1, 4);
    \coordinate (actE) at (0.1, 0, 2);
    \coordinate (actF) at (0.1, 1, 2);
    \coordinate (actG) at (0, 1, 2);

    \draw[act] (actA) -- (actB) -- (actC) -- (actD) -- (actA);
    \draw[act] (actF) --  (actG) -- (actD) -- (actC) -- (actF);
    \node[] at (-0.5, 0.3, 4) {PMC};
    \node[] at (-0.5, 2, 4) {Setup 2};

  \end{scope}
  % active region
  \begin{scope}[shift={(0.1, 1.1, 0)}]
    \coordinate (actA) at (0, 0, 4);
    \coordinate (actB) at (6, 0, 4);
    \coordinate (actC) at (6, 1, 4);
    \coordinate (actD) at (0, 1, 4);
    \coordinate (actE) at (6, 0, 2);
    \coordinate (actF) at (6, 1, 2);
    \coordinate (actG) at (0, 1, 2);

    \draw[act] (actA) -- (actB) -- (actC) -- (actD) -- (actA);
    \draw[act] (actF) --  (actG) -- (actD) -- (actC) -- (actF);
    \node[] at (1.2, 0.3, 4) {Active region};
  \end{scope}
  % PML
  \begin{scope}[shift={(6.1, 1.1, 0)}]
    \coordinate (actA) at (0, 0, 4);
    \coordinate (actB) at (0.1, 0, 4);
    \coordinate (actC) at (0.1, 1, 4);
    \coordinate (actD) at (0, 1, 4);
    \coordinate (actE) at (0.1, 0, 2);
    \coordinate (actF) at (0.1, 1, 2);
    \coordinate (actG) at (0, 1, 2);

    \draw[act] (actA) -- (actB) -- (actC) -- (actD) -- (actA);
    \draw[act] (actF) --  (actG) -- (actD) -- (actC) -- (actF);
    \draw[act] (actB) -- (actC) --(actF) -- (actE) -- (actB);
    \draw[ld] (-6, -0.1, 4) -- (0, -0.1, 4);
    \node[] at (-4.6, -0.4, 4) {$L_2 = \SI{10}{\milli\meter}$};
    \begin{scope}[canvas is yz plane at x=0.1, transform shape]
      \node[rotate=-90] at (0.5,3) {PMC};
    \end{scope}
  \end{scope}

  %% Setup 2
  \begin{scope}[shift={(0, 3, 0)}]
    \begin{scope}[shift={(0, 1.1, 0)}]
      \coordinate (actA) at (0, 0, 4);
      \coordinate (actB) at (0.1, 0, 4);
      \coordinate (actC) at (0.1, 1, 4);
      \coordinate (actD) at (0, 1, 4);
      \coordinate (actE) at (0.1, 0, 2);
      \coordinate (actF) at (0.1, 1, 2);
      \coordinate (actG) at (0, 1, 2);

      \draw[act] (actA) -- (actB) -- (actC) -- (actD) -- (actA);
      \draw[act] (actF) --  (actG) -- (actD) -- (actC) -- (actF);
      \node[] at (-0.5, 0.3, 4) {PMC};
      \node[] at (-0.5, 2, 4) {Setup 1};
    \end{scope}
    % active region
    \begin{scope}[shift={(0.1, 1.1, 0)}]
      \coordinate (actA) at (0, 0, 4);
      \coordinate (actB) at (3, 0, 4);
      \coordinate (actC) at (3, 1, 4);
      \coordinate (actD) at (0, 1, 4);
      \coordinate (actE) at (3, 0, 2);
      \coordinate (actF) at (3, 1, 2);
      \coordinate (actG) at (0, 1, 2);

      \draw[act] (actA) -- (actB) -- (actC) -- (actD) -- (actA);
      \draw[act] (actF) --  (actG) -- (actD) -- (actC) -- (actF);
      \node[] at (1.2, 0.3, 4) {Active region};
    \end{scope}
    % PML
    \begin{scope}[shift={(3.1, 1.1, 0)}]
      \coordinate (actA) at (0, 0, 4);
      \coordinate (actB) at (1, 0, 4);
      \coordinate (actC) at (1, 1, 4);
      \coordinate (actD) at (0, 1, 4);
      \coordinate (actE) at (1, 0, 2);
      \coordinate (actF) at (1, 1, 2);
      \coordinate (actG) at (0, 1, 2);

      \draw[act] (actA) -- (actB) -- (actC) -- (actD) -- (actA);
      \draw[act] (actF) --  (actG) -- (actD) -- (actC) -- (actF);
      \draw[act] (actB) -- (actC) --(actF) -- (actE) -- (actB);
      \node[] at (0.5, 0.3, 4) {PML};
      \draw[ld] (-3, -0.1, 4) -- (0, -0.1, 4);
      \node[] at (-1.6, -0.4, 4) {$L_1 = \SI{5}{\milli\meter}$};
      \draw[dashed] (-0.6,-3.5,4) -- (-0.6, 2, 4)  -- (-0.6, 2, 2) node[above] {Position} -- (-0.6, -3.5, 2) --(-0.6, -3.5, 4);
    \end{scope}
    %PMC
    \begin{scope}[shift={(4.1, 1.1, 0)}]
      \coordinate (actA) at (0, 0, 4);
      \coordinate (actB) at (0.1, 0, 4);
      \coordinate (actC) at (0.1, 1, 4);
      \coordinate (actD) at (0, 1, 4);
      \coordinate (actE) at (0.1, 0, 2);
      \coordinate (actF) at (0.1, 1, 2);
      \coordinate (actG) at (0, 1, 2);

      \draw[act] (actA) -- (actB) -- (actC) -- (actD) -- (actA);
      \draw[act] (actF) --  (actG) -- (actD) -- (actC) -- (actF);
      \draw[act] (actB) -- (actC) --(actF) -- (actE) -- (actB);
      \begin{scope}[canvas is yz plane at x=0.1, transform shape]
        \node[rotate=-90] at (0.5,3) {PMC};
      \end{scope}
    \end{scope}
  \end{scope}
\end{tikzpicture}
  \caption{Illustration of the two different simulation setups under numerical investigation. Setup 1 describes a THz gain medium with length $L_1=\SI{5}{\milli\meter}$ truncated with a PML boundary containing \num{200} gridpoints terminated by a PMC layer, setup 2 is based on the same material system with twice the length and two PMC layers at the facets.}
  \label{fig:pml_setup}
\end{figure}
We validate the numerical stability of our modified PML model by
investigating the light propagation in the aforementioned THz QCL gain
structure for two different simulation setups, as depicted in
Fig.~\ref{fig:pml_setup}. Setup 1 consists of a an active region of length
\SI{5}{\milli\meter}, which is terminated by a perfect magnetic conductor
(PMC) layer on the left and a \num{200}-layer PML region on the right facet.
Setup 2 has a simulation domain twice as long as in setup 1 and is terminated
with a PMC boundary layer on both sides.

As pointed out in~\cite{Seitner_2021}, the chromatic dispersion in the
investigated THz structure arises not only from the quantum system, but also
from the background material and the waveguide itself. Here, the chromatic
dispersion introduced by the background system is neglected, as we are more
interested in the interaction of the PML region with the quantum system. In
order to extract the spectral gain profile we excite the system with a
Gaussian field pulse
$\function{E_z}{0,t}=A\exp\left[-\left(t-t_0\right)/\tau\right]^2\sin{\left(2\pi
    f_0t\right)}$ at the left facet of setup 1 and measure the amplified field at
the position $x=L=\SI{4}{\milli\meter}$, indicated by a dashed rectangle in
Fig.~\ref{fig:pml_setup}. The pulse parameters are
$A=\SI{1e-3}{\volt\per\meter}$, $\tau=\SI{0.707}{\pico\second}$,
$t_0=\SI{30}{\pico\second}$, and $f_0=\SI{3.8}{\tera\hertz}$. The amplitude
gain coefficient can be calculated from the Fourier transforms of the
recorded electric fields $\function{E_\mathrm{in}}{\omega}$ and
$\function{E_\mathrm{out}}{\omega}$ by
\begin{equation}
  \function{g}{\omega}=\frac{1}{L}\ln{\left(\frac{\abs{\function{E_\mathrm{out}}{\omega}}}{\abs{\function{E_\mathrm{in}}{\omega}}}\right)}\, .
\end{equation}
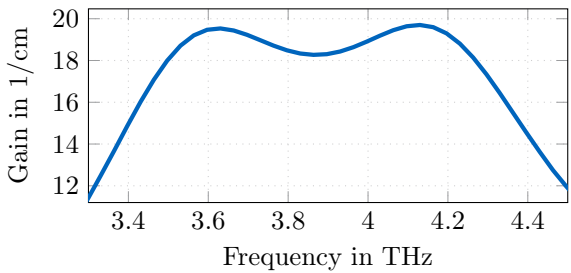
\begin{figure}[b]
  \centering
  \begin{tikzpicture}

  \pgfplotstableread{freqs	gain
    2.997565e+12	-8.002916e+02
    3.030871e+12	-5.515209e+02
    3.064177e+12	-5.202227e+02
    3.097484e+12	-5.203616e+02
    3.130790e+12	-3.985414e+02
    3.164096e+12	-3.511447e+02
    3.197402e+12	-2.681508e+02
    3.230709e+12	-1.684801e+02
    3.264015e+12	-7.693185e+01
    3.297321e+12	3.351725e+01
    3.330627e+12	1.486829e+02
    3.363934e+12	2.661666e+02
    3.397240e+12	3.864425e+02
    3.430546e+12	5.022064e+02
    3.463853e+12	6.081439e+02
    3.497159e+12	6.994308e+02
    3.530465e+12	7.711102e+02
    3.563771e+12	8.203340e+02
    3.597078e+12	8.470261e+02
    3.630384e+12	8.534321e+02
    3.663690e+12	8.436468e+02
    3.696996e+12	8.230189e+02
    3.730303e+12	7.971500e+02
    3.763609e+12	7.710086e+02
    3.796915e+12	7.487198e+02
    3.830222e+12	7.334169e+02
    3.863528e+12	7.269843e+02
    3.896834e+12	7.302361e+02
    3.930140e+12	7.429818e+02
    3.963447e+12	7.638616e+02
    3.996753e+12	7.903842e+02
    4.030059e+12	8.190005e+02
    4.063365e+12	8.452139e+02
    4.096672e+12	8.639082e+02
    4.129978e+12	8.700413e+02
    4.163284e+12	8.595189e+02
    4.196591e+12	8.298821e+02
    4.229897e+12	7.807611e+02
    4.263203e+12	7.141069e+02
    4.296509e+12	6.337237e+02
    4.329816e+12	5.439214e+02
    4.363122e+12	4.491939e+02
    4.396428e+12	3.547253e+02
    4.429734e+12	2.628426e+02
    4.463041e+12	1.747826e+02
    4.496347e+12	9.709062e+01
    4.529653e+12	2.428904e+01
    4.562960e+12	-4.425152e+01
    4.596266e+12	-9.095257e+01
    4.629572e+12	-1.478254e+02
    4.662878e+12	-1.954855e+02
    4.696185e+12	-2.024060e+02
    4.729491e+12	-2.711320e+02
    4.762797e+12	-2.824056e+02
    4.796104e+12	-2.466961e+02
    4.829410e+12	-4.005562e+02
    4.862716e+12	-2.976020e+02
    4.896022e+12	-2.269219e+02
    4.929329e+12	-6.294290e+02
    4.962635e+12	-2.001596e+02
    4.995941e+12	-1.110645e+02
  }\datatable

  \begin{axis}[%
    width = 0.95\columnwidth,
    height = 0.5\columnwidth,
    xlabel={Frequency in $\si{\tera\hertz}$},
    xmax=4.5,
    xmin = 3.3,
    scaled ticks=false,
    ylabel={Gain in $\si{\per\centi\meter}$}, grid=both, grid
    style={draw=gray!50, dotted} ] \addplot[TUMBlue,ultra thick] table[x
      expr=1e-12*\thisrow{freqs}, y expr= 1e-2*\thisrow{gain}+11]{\datatable};

  \end{axis}

\end{tikzpicture}%
  \caption{Simulated spectral gain profile of the THz QCL gain structure.}
  \label{fig:gain_spec}
\end{figure}
The resulting spectral gain profile of the THz QCL gain medium is presented
in Fig.~\ref{fig:gain_spec}. We observe two gain peaks at
$f=\SI{3.63}{\tera\hertz}$ and $f=\SI{4.13}{\tera\hertz}$ which correspond to
the optical transitions $\mathrm{UL}\rightarrow\mathrm{LL2}$ and
$\mathrm{UL}\rightarrow\mathrm{LL1}$, respectively. Our simulation yields a
peak gain of $g_\mathrm{p}=\SI[per-mode=power]{19.7}{\per\centi\meter}$.

\begin{figure}[ptb]
  \centering
  \input{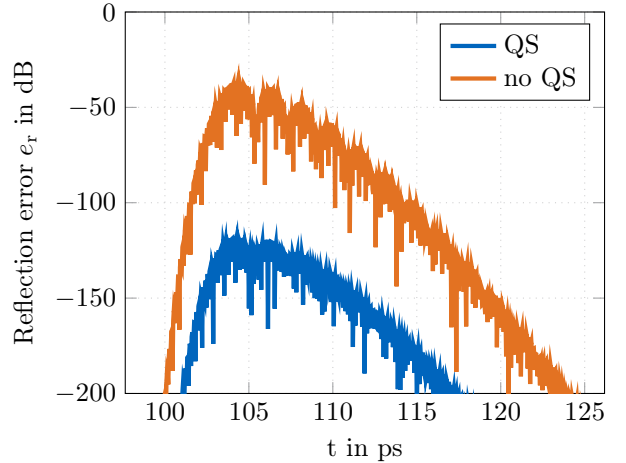}
  \caption{Reflection error for two PML configurations. The blue curve represents the reflection error including the macroscopic polarization of the QS in the PML region, whereas the orange curve represents the results obtained from the original PML formulation introduced by Gedney~\cite{gedney1996anisotropic1,gedney1996anisotropic2}.}
  \label{fig:refl_err}
\end{figure}

We now quantify the reflection behavior at the PML boundary by introducing
the reflection error $e_\mathrm{r}$, computed as
\begin{equation}
  \function{e_\mathrm{r}}{t}=\left.\frac{\abs{\function{E_1}{t}-\function{E_2}{t}}}{\max{\left(\abs{\function{E_2}{t}}\right)}}\right|_{x=\SI{4}{\milli\meter}}\,,
\end{equation}
with $\function{E_1}{t}$ being the reflected electric field in setup 1. The
field $\function{E_2}{t}$ from setup 2 is used as reference value. As already
pointed out, we also include the quantum system of the active region in the
adjacent PML region to obtain better impedance matching. The amplitude gain
displayed in Fig.~\ref{fig:gain_spec} now also acts on the outgoing electric
field in the PML. However, the artificially introduced losses from
Eq.~\eqref{eq:sigma_x} outperform the unintended amplification by far and the
light gets efficiently absorbed in the PML region. In
Fig.~\ref{fig:refl_err}, we compare the results of our implementation to a
similar PML region, but here we drop the quantum system (QS) in the boundary.
Due to the resulting impedance mismatch such a scenario yields a maximum
reflection error of $e_\mathrm{r}\thicksim\SI{-36}{\decibel}$ for the
exemplary THz QCL setup. By taking into account the QS in the PML layers, we
obtain a reflection error $e_\mathrm{r}$ of less than $\SI{-118}{\decibel}$.

\section{Conclusion}

In this paper, we presented an extension to the existing generalized open
source Maxwell-Bloch equation framework with the incorporation of perfectly
matched layer absorbing boundary conditions. By applying the new
implementation to a THz QCL gain medium an efficient attenuation of outwards
propagating fields has been demonstrated. We integrate the quantum gain
system into the PML region and thus assure an improved outcoupling of the
optical field. A significantly reduced reflection error as compared to
previous PML models has been achieved.

\section{Acknowledgements}

The authors acknowledge financial support by the European Union's Horizon
2020 research and innovation programme under grant agreement No 820419 --
Qombs Project ''Quantum simulation and entanglement engineering in quantum
cascade laser frequency combs'' (FET Flagship on Quantum Technologies).

\bibliographystyle{IEEEtran}
\bibliography{references}

\end{document}